\documentclass[12pt,english]{paper}
\usepackage{amsthm,amsmath,amssymb}
\numberwithin{equation}{section}
\usepackage[T1]{fontenc}
\usepackage[latin1]{inputenc}
\usepackage{graphicx}
\usepackage{cite}
\usepackage{slashed}
\makeatletter

\newcommand{\bibstyle@aas}{\bibpunct{(}{)}{;}{a}{}{,}}

\makeatletter

\makeatletter

\usepackage{geometry}

\makeatletter

\usepackage{epsfig}

\makeatother

\makeatother

\makeatother

\usepackage{babel}
\makeatother
\begin{document}

\title{Measurement of Milli-Charged Particles with a running electromagnetic coupling constant at IceCube}

\author{Ye Xu}

\maketitle

\begin{flushleft}
School of Electronic, Electrical Engineering and Physics,  Fujian University of Technology, Fuzhou 350118, China
\par
e-mail address: xuy@fjut.edu.cn
\end{flushleft}

\begin{abstract}
It is postulated that heavy dark matter $\phi$ with a mass on the order of TeV, once captured by the Earth, can decay into relativistic milli-charged particles (MCPs). These MCPs are potentially detectable at the IceCube neutrino telescope. In this study, MCPs are modeled within the massless hidden photon framework, where they interact with nuclei via a running electromagnetic coupling constant, thereby enabling the prediction of their expected event rates and fluxes at IceCube. The expected number of background neutrino events at IceCube has also been evaluated. Under the assumption that no signal events are observed over a 10-year period at IceCube, upper limits on the MCP flux have been derived at the 90\% confidence level. The results suggest that MCPs originating from the Earth's core could be directly detected at IceCube at energies around $\mathcal{O}(1\ \text{TeV})$ for a fractional charge-squared range of $5.65\times10^{-5} \lesssim \epsilon^2 \lesssim 1.295\times10^{-3}$. Furthermore, with 10 years of IceCube data, a new region in the $m_{\text{MCP}}$-$\epsilon$ parameter space--specifically, $4\ \text{GeV} < m_{\text{MCP}} < 100\ \text{GeV}$ and $5.51\times10^{-2} < \epsilon < 0.612$--has been excluded.
\end{abstract}

\section{Introduction}
According to the Planck data from cosmic microwave background measurements, approximately 26\% of the total energy density of the Universe is composed of non-baryonic dark matter (DM)\cite{Planck2016}. However, experimental constraints on dark matter have become increasingly stringent, ruling out most of its parameter space\cite{XENON1T,PANDAX,fermi,antares-icecube-dm-MW,icecubedm-sun,antaresdm-sun,CAST,GlueX,NGC1275,Chooz,dayabayMINOS}. A small fraction of DM may consist of relativistic particles in the Universe. Such particles could reach the IceCube detector and be observed at energies on the order of TeV.
\par
The origin of dark matter (DM) remains unknown, despite a wealth of compelling observational evidence\cite{BHS}. In the scenario considered here, there exist at least two distinct species of DM in the Universe. One is a heavy DM particle with a mass on the order of TeV, while the other is a light DM particle. The heavy particle ($\phi$) is a thermal relic produced in the early Universe and constitutes the majority of present-day DM. The other component is a stable light fermion ($\chi$), which arises from the decay process $\phi \to \chi\bar{\chi}$, analogous to the DM decay channel discussed in Ref. \cite{FR}. This light fermion is a milli-charged particle (MCP)---an alternative DM candidate \cite{GH,CY,FLN}---carrying a small electric charge $\epsilon e$, where $e$ is the electron charge and $\epsilon < 1$.
\par
Searches for MCPs have been conducted across various fronts, including cosmological and astrophysical observations, accelerator experiments, studies of ortho-positronium decay and Lamb shifts, and direct DM detection experiments. These efforts have placed stringent constraints on the parameter $\epsilon$ \cite{CM,DHR,GH,DCB,VR,SLAC,Xenon,LS,OP,Lamb,SUN}. In this model, $m_\chi \ll m_\phi$, with $m_\phi \geq 10\ \mathrm{TeV}$, and the MCP mass is taken to be below 100 GeV. Due to the decay of the long-lived $\phi$ -- with a lifetime $\tau_\phi > t_0$ (where $t_0 \sim 10^{17}\ \mathrm{s}$ is the age of the Universe, and here $\tau_\phi \geq 10^{18}\ \mathrm{s}$) \cite{AMO,EIP} -- the present-day DM may also include a very minor component of MCPs. These MCPs possess energies approximately equal to $m_\phi/2$, where $m_\phi$ denotes the mass of $\phi$. Furthermore, it is assumed that $\phi$ decays exclusively via $\phi \to \chi\bar{\chi}$.
\par
As $\phi$ particles from the Galactic halo sweep through the Earth, they can be captured by it. In this work, we assume that MCPs interact with nuclei via a running electromagnetic coupling constant. A key concept in quantum field theory is that coupling constants--which quantify the strength of an interaction--are not truly constant but depend on the energy scale of the process. In quantum electrodynamics (QED), the fine-structure constant $\alpha$ runs with the squared four-momentum transfer $Q^2$. This running behavior arises from contributions due to lepton and quark loops \cite{EMN}.
\par
\begin{figure}
 \centering
 \includegraphics[width=0.9\textwidth]{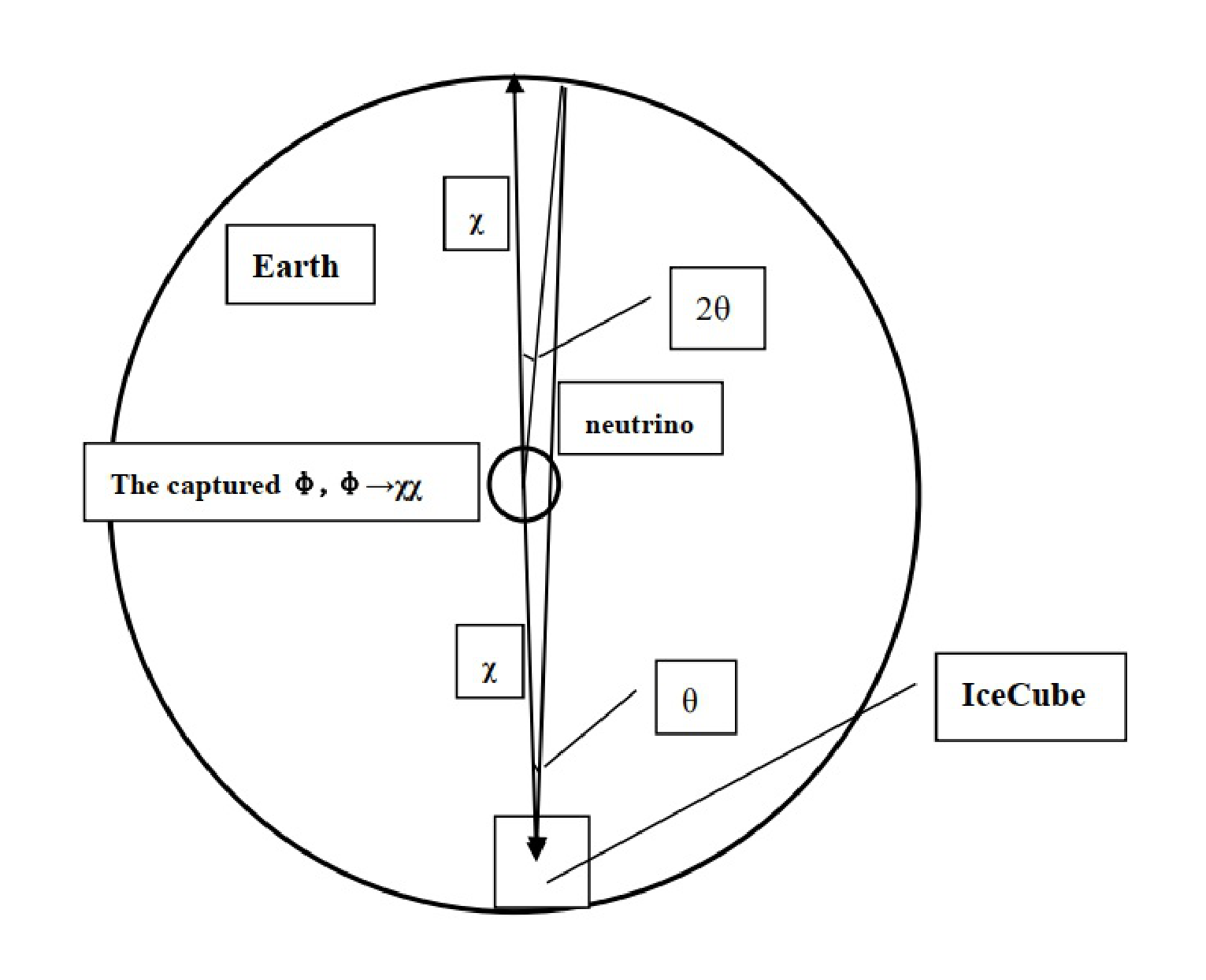}
 \caption{Schematic of MCPs produced from decay of heavy DM particles captured in the Earth's core, which then traverse the Earth and ice to become detectable by the detector like IceCube neutrino telescope}
 \label{fig:fig}
\end{figure}

Only high-energy MCPs originating from the Earth's core can be detected in this scenario (see Fig.~\ref{fig:fig}). These particles can be directly observed by the IceCube neutrino telescope through deep inelastic scattering (DIS) with ice nuclei after traversing the Earth and the ice sheet. The detectability of such signals will also be discussed in this work. The relevant measurement is subject to a background dominated by atmospheric neutrinos--produced in cosmic-ray interactions in the Earth's atmosphere--as well as astrophysical neutrinos.
\section{Flux of MCPs which reach the earth surface}
As the Galactic halo $\phi$ particles sweep through the Earth, they can collide with atomic nuclei and become gravitationally captured. Those accumulated in the Earth subsequently decay into MCPs at a significant rate. The number of captured $\phi$ particles is governed by the evolution equation from Ref. \cite{BCH}:
\begin{equation}
\frac{dN}{dt} = C_{\text{cap}} - C_{\text{ann}} N^2 - C_{\text{evp}} N
\end{equation}
where $C_{\text{cap}}$, $C_{\text{ann}}$, and $C_{\text{evp}}$ represent the capture, annihilation, and evaporation rates, respectively. The evaporation rate is only relevant for dark matter masses below 5 GeV \cite{BCH}, which is much lower than the mass scale considered here ($m_\phi \sim \mathcal{O}(\text{TeV})$). Therefore, evaporation contributes negligibly to the accumulation of $\phi$ in the Earth within this work.

The annihilation rate is given by $\Gamma_{\text{ann}} = \displaystyle\frac{1}{2} C_{\text{ann}} N^2$, and following Refs. \cite{BCH, FST}, it can be expressed as:
\begin{equation}
\Gamma_{\text{ann}} = \frac{C_{\text{cap}}}{2} \tanh^2\left( \frac{t}{\tau} \right) \approx \frac{C_{\text{cap}}}{2}, \quad \text{for} \quad t \gg \tau
\end{equation}
where $\tau = (C_{\text{cap}} C_{\text{ann}})^{-1/2}$ is the characteristic timescale set by the balance between capture and annihilation. For the Earth, the approximation $\tanh^2(t/\tau) \approx 1$ holds at late times $t \gg \tau$ \cite{BCH}.

\begin{figure}
 \centering
 \includegraphics[width=0.9\textwidth]{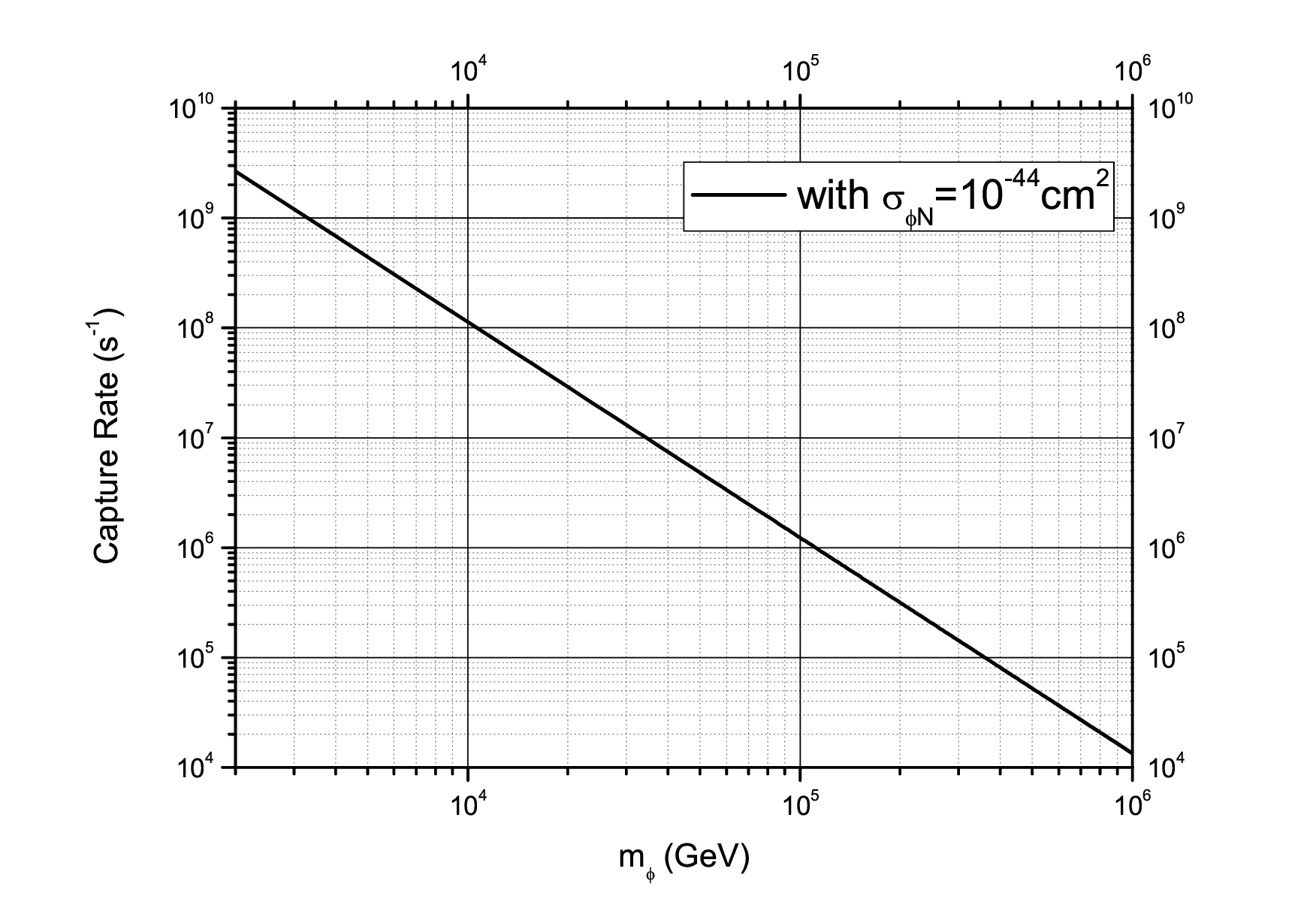}
 \caption{The capture rates of $\phi$'s. This figure shows these rates at which $\phi$'s are captured to the interior of the Earth, for a scattering cross section of $\sigma_{\phi N}=10^{-44}$ cm$^2$.}
 \label{fig:cap}
\end{figure}

The capture rate $C_{\text{cap}}$ is proportional to $\sigma_{\phi N} / m_\phi$ \cite{BCH, JKK}, where $\sigma_{\phi N}$ is the scattering cross section between $\phi$ particles and nuclei. In this calculation, only the spin-independent cross section is considered, and $\sigma_{\phi N}$ is taken as $10^{-44}~\text{cm}^2$ for $m_\phi \sim \mathcal{O}(\text{TeV})$ \cite{XENON1T, PANDAX}. Furthermore, as shown in Ref. \cite{FST}, the captured $\phi$ particles are concentrated around the Earth's center. Figure~\ref{fig:cap} shows the corresponding capture rates for $\sigma_{\phi N} = 10^{-44}~\text{cm}^2$, computed according to the method in Ref. \cite{SE}.

The MCPs reaching the IceCube detector are produced from the decay of $\phi$ particles in the Earth's core. These MCPs must then traverse the Earth and may interact with nuclei along their path. The number $N_e$ of MCPs produced in the core is given by:
\begin{equation}
\begin{aligned}
N_e &= 2N_0 \left( \exp\left( -\frac{t_0}{\tau_\phi} \right) - \exp\left( -\frac{t_0 + T}{\tau_\phi} \right) \right), \quad \text{with} \quad T \ll \tau_\phi\\
&\approx 2N_0 \frac{T}{\tau_\phi} \exp\left( -\frac{t_0}{\tau_\phi} \right)
\end{aligned}
\end{equation}
where $N_0 = \int_0^{t_e} \displaystyle\frac{dN}{dt} dt$ is the total number of $\phi$ particles captured in the Earth, $t_e$ and $t_0$ are the ages of the Earth and the Universe, respectively, and $T = 10$ years is the data-taking period of IceCube. The Earth's radius is denoted by $R_e$.

The flux $\Phi_{\text{MCP}}$ of MCPs arriving at the IceCube detector from the Earth's core is then described by:
\begin{equation}
\Phi_{\text{MCP}} = \frac{N_e}{4\pi R_e^2} \frac{d}{dE} \left[ \exp\left( -\frac{R_e}{L_e^\chi} \right) \right]
\end{equation}
where $L_e^\chi$ is the interaction length of MCPs within the Earth.
\section{MCP and neutrino interactions with nuclei}
The hidden photon model introduces a second unbroken "mirror" U(1)$^{\prime}$ symmetry. The associated massless hidden photon field can undergo kinetic mixing with the Standard Model (SM) photon, enabling MCPs charged under U(1)$^{\prime}$ to acquire a small effective coupling to the SM photon \cite{Holdom}. The parameter $\epsilon$ quantifies this kinetic mixing between the two photons. In this model, MCPs interact with nuclei via a neutral current (NC) interaction mediated by the effective photon arising from the kinetic mixing of the SM and hidden photons.
\par
A well-motivated interaction, permitted by SM symmetries and providing a "portal" between SM particles and MCPs, is given by $\displaystyle\frac{\epsilon}{2}F_{\mu\nu}F^{\prime\mu\nu}$. The corresponding interaction Lagrangian is:

\begin{equation}
\mathcal{L} =\sum_qe_q\bar{q}\gamma^{\mu}qA_{\mu} -\frac{1}{4}F^{\prime}_{\mu\nu}F^{\prime\mu\nu}+\bar{\chi}(i\slashed{D}-m_{\chi})\chi-\frac{\epsilon}{2}F_{\mu\nu}F^{\prime\mu\nu}
\end{equation}
Here, the sum runs over quark flavors in the nucleon, $e_q$ is the electric charge of quark $q$, $A_{\mu}$ is the SM photon vector potential, and $F_{\mu\nu}$ and $F^{{\prime}{\mu\nu}}$ are the field strength tensors of the SM and hidden photons, respectively. $m_{\chi}$ denotes the MCP mass.
\par
The DIS cross section of MCPs on nuclei is computed using the same model described in Section 3 of my previous work \cite{SUN}. Since the effective coupling between MCPs and the mediator is $\epsilon^2 \alpha$, the DIS cross section for MCP-nucleus scattering equals $\epsilon^2$ times the corresponding NC DIS cross section for electrons on nuclei via photon exchange, i.e.,

\begin{equation}
\sigma_{\chi N} \approx \epsilon^2 \sigma^{\gamma}_{eN}
\end{equation}
where $\chi$ represents an MCP with charge $\epsilon e$, $N$ is a nucleon, and $\sigma^{\gamma}_{eN}$ is the photon-mediated electron-nucleon DIS cross section.

The running electromagnetic coupling $\alpha(Q^2)$ is parameterized as \cite{EMN}:

\begin{equation}
\alpha(Q^2) = \frac{\alpha_0}{1 - \Delta\alpha(Q^2)}
\end{equation}
Here, $\alpha_0$ is the fine-structure constant, and $\Delta\alpha(Q^2)$ encodes the running of $\alpha$, receiving contributions from lepton loops, top-quark loops, and loops from the five light quark flavors. Values for $\alpha(Q^2)$ can be derived from measurements of the running coupling at LEP \cite{S.Mele}.
\par
In the energy range 1 TeV - 1 PeV, the total DIS cross section for MCPs on nuclei can be approximated by a simple power law:

\begin{equation}
\sigma_{\chi N} \approx 3.176 \times 10^{-31} \epsilon^2 \left( \frac{E_{\chi}}{1 \text{GeV}} \right)^{0.156} \text{cm}^2
\end{equation}
where $E_{\chi}$ is the MCP energy.
\par
For neutrino-nucleus DIS cross sections in the lab frame, simple power-law approximations are used for neutrino energies above 1 TeV \cite{BHM}:

\begin{equation}
\sigma_{\nu N}(\text{CC}) = 4.74 \times 10^{-35} \left( \frac{E_{\nu}}{1 \text{GeV}} \right)^{0.251} \text{cm}^2
\end{equation}

\begin{equation}
\sigma_{\nu N}(\text{NC}) = 1.80 \times 10^{-35} \left( \frac{E_{\nu}}{1 \text{GeV}} \right)^{0.256} \text{cm}^2
\end{equation}
Here, $\sigma_{\nu N}(\text{CC})$ and $\sigma_{\nu N}(\text{NC})$ are the DIS cross sections for neutrino-nucleus scattering via charged-current and neutral-current interactions, respectively, and $E_{\nu}$ is the neutrino energy.
\par
\begin{figure}
 \centering
 \includegraphics[width=0.9\textwidth]{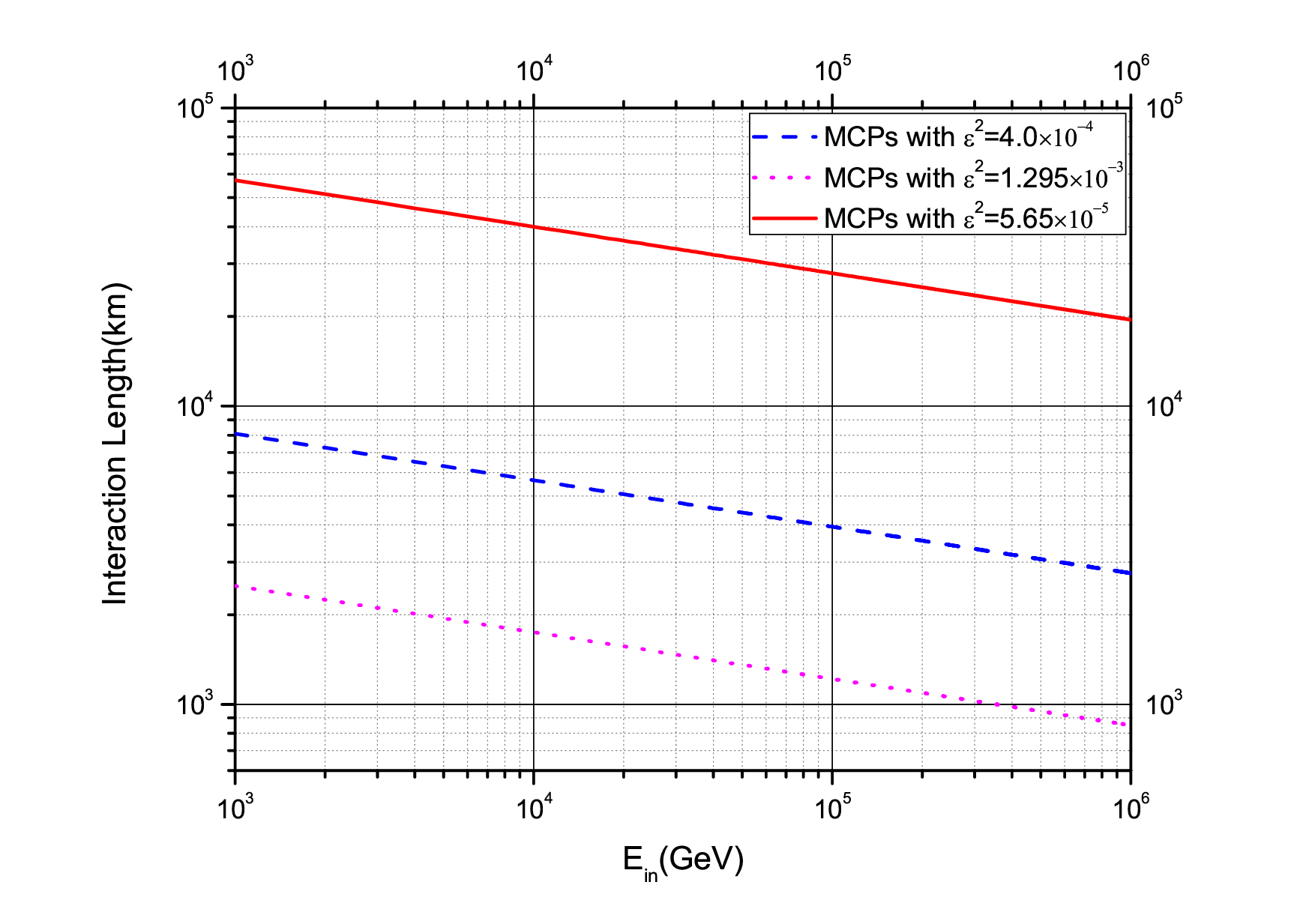}
 \caption{The MCP interaction lengths with the Earth were computed when $\epsilon^2$ = $5.65\times10^{-5}$(red solid line), $4\times10^{-4}$(blue dash line) and $1.295\times10^{-3}$(magenta dot line), respectively.}
 \label{fig:L_dm}
\end{figure}

The interaction lengths for MCPs and neutrinos are given by:

\begin{equation}
L^{\nu,\chi}=\frac{1}{N_A\sigma_{\nu,\chi N}\rho}
\end{equation}
where $N_A = 6.022 \times 10^{23}~\text{cm}^{-3}$ (water equivalent) is the Avogadro constant, and $\rho = \rho_m / \rho_w$, with $\rho_m$ being the density of the material through which MCPs or neutrinos propagate, and $\rho_w$ the density of water. Figure~\ref{fig:L_dm} shows the MCP interaction lengths in the Earth for $\epsilon^2 = 5.65 \times 10^{-5}$ (red solid line), $4 \times 10^{-4}$ (blue dashed line), and $1.295 \times 10^{-3}$ (magenta dotted line).
\par
Since the energy deposited by MCPs is proportional to $q^2$ (where $q$ is the electric charge of the MCP), an MCP with charge $\epsilon e$ exhibits an energy loss reduced by a factor of $\epsilon^2$ compared to a particle with charge $1e$. For instance, if $\epsilon = 0.1$, the energy loss of an MCP is 0.01 times that of an electron traversing the same medium. In this analysis, energy loss of MCPs during propagation through the Earth is neglected.
\section{Evaluation of the numbers of expected MCPs and neutrinos at IceCube}
The IceCube detector is deployed in the deep ice below the geographic South Pole \cite{icecube2004}. When high-energy MCPs pass through the IceCube detector, they interact with atomic nuclei in the ice through deep-inelastic scattering (DIS). This interaction mechanism is analogous to the neutral-current (NC) DIS of neutrinos with nuclei, in which secondary particles develop into detectable cascade events.
\par
The reconstruction of cascade directions from the topological distribution of Cherenkov photon signals allows for directional discrimination. Consequently, neutrinos misidentified as MCPs must originate from the same celestial region as the expected MCP signal, rather than from arbitrary directions. This implies that the relevant neutrino background reaching the IceCube detector after traversing the Earth is restricted to events arriving from the vicinity of the geographic North Pole. To accommodate reconstruction uncertainties, signal selection is performed using windows defined by the uncertainties in physical parameters. This selection criterion reduces the number of detectable MCP signals, while the background event count is determined by integrating over these same windows. In the present analysis, only the uncertainties in energy and directional reconstruction are considered when evaluating the expected numbers of MCP and neutrino events.
\par
The expected number of detected MCPs, N$_{det}$, follows the relation:
\begin{center}
\begin{equation}
\frac{dN_{det}}{dE} =C_1\times C_2\times A_{eff}(E)\Phi_{MCP}P(E)
\end{equation}
\end{center}
where $A_{eff}(E)$ denotes the energy-dependent effective detection area of IceCube, obtained from the Fig. 3 in Ref.\cite{icecube2014a}, and E represents the energy of the incident particle. The factor $C_1$ = 68.3\% corresponds to the fraction of reconstructed MCP events falling within a one-standard-deviation energy uncertainty window, while $C_2$ = 50\% represents the fraction within a one-median-angular-uncertainty window. $P(E)$ is the probability that the MCPs reaching the IceCube detector can be detected by IceCube. The detection probability P(E) is expressed as:
\begin{center}
\begin{equation}
P(E)=1-exp(-\displaystyle\frac{D}{L^{\chi}_{ice}}).
\end{equation}
\end{center}
where $L^{\chi}_{ice}$ is the MCP interaction length in ice. D = 1 km represents the effective length in the IceCube detector adopted in this work.
\par
Following the rejection of track-like events, the remaining background comprises both astrophysical and atmospheric neutrinos traversing the IceCube detector. For muon neutrinos, only NC interactions are considered. The astrophysical neutrino flux is parameterized as \cite{icrc2023}:
\begin{center}
\begin{equation}
\Phi_{\nu}^{astro}=\Phi_{astro}\times\left(\displaystyle\frac{E_{\nu}}{100TeV}\right)^{-(\alpha+\beta log_{10}(\frac{E_{\nu}}{100TeV}))}\times10^{-18}GeV^{-1} cm^{-2}s^{-1}sr^{-1}
\end{equation}
\end{center}
where $\Phi_{\nu}^{astro}$ denotes the total astrophysical neutrino flux, and the parameters $\Phi_{astro}$, $\alpha$ and $\beta$ are provided in Table 2 of Ref.\cite{icrc2023}. The atmospheric neutrino flux follows the formulation\cite{SMS}:
\begin{center}
\begin{equation}
\Phi_{\nu}^{atm} = C_{\nu}\left(\displaystyle\frac{E_{\nu}}{1GeV}\right)^{-(\gamma_0+\gamma_1x+\gamma_2x^2)}GeV^{-1} cm^{-2}s^{-1}sr^{-1}
\end{equation}
\end{center}
with $x=log_{10}(E_{\nu}/1GeV)$. Here, $\Phi_{\nu}^{atm}$ represents the atmospheric neutrino flux, and the coefficients $C_{\nu}$ ($\gamma_0$, $\gamma_1$ and $\gamma_2$) are given in Table III of Ref.\cite{SMS}.
\par
Neutrinos falling within the defined energy and angular windows are considered signal candidates. The expected number of neutrino events, N$_{\nu}$, is evaluated through the integral:
\begin{center}
\begin{equation}
\frac{dN_{\nu}}{dE} = \int_T \int_{\theta_{min}}^{\theta_{max}} A_{eff}(E)(\Phi_{\nu}^{astro}+\Phi_{\nu}^{atm}) P(E,\theta)\frac{2\pi R_e^2 sin2\theta}{D_e(\theta)^2} d\theta dt
\end{equation}
\end{center}
where $D_e(\theta)=2R_e cos\theta$ is the distance traveled through the Earth, $\theta$ represents the zenith angle at IceCube (see Fig.~\ref{fig:fig}), $\theta_{min}$ = 0 and $\theta_{max}$ = $\sigma_{\theta}$, with $\sigma_{\theta}$ denoting the median angular uncertainty for cascade events in IceCube. The standard energy and median angular uncertainties are obtained from the Ref.\cite{icecube2021ICRC} and Ref.\cite{icecube2013}, respectively. The probability factor $P(E,\theta)$ is defined as:
\begin{center}
\begin{equation}
P(E,\theta)=exp(-\displaystyle\frac{D_e(\theta)}{L^{\nu}_{e}})\left(1-exp(-\displaystyle\frac{D}{L^{\nu}_{ice}})\right)
\end{equation}
\end{center}
where $L^{\nu}_{e,ice}$ represent the neutrino interaction length in the Earth and ice, respectively.
\section{Results}
\begin{figure}
 \centering
 \includegraphics[width=0.9\textwidth]{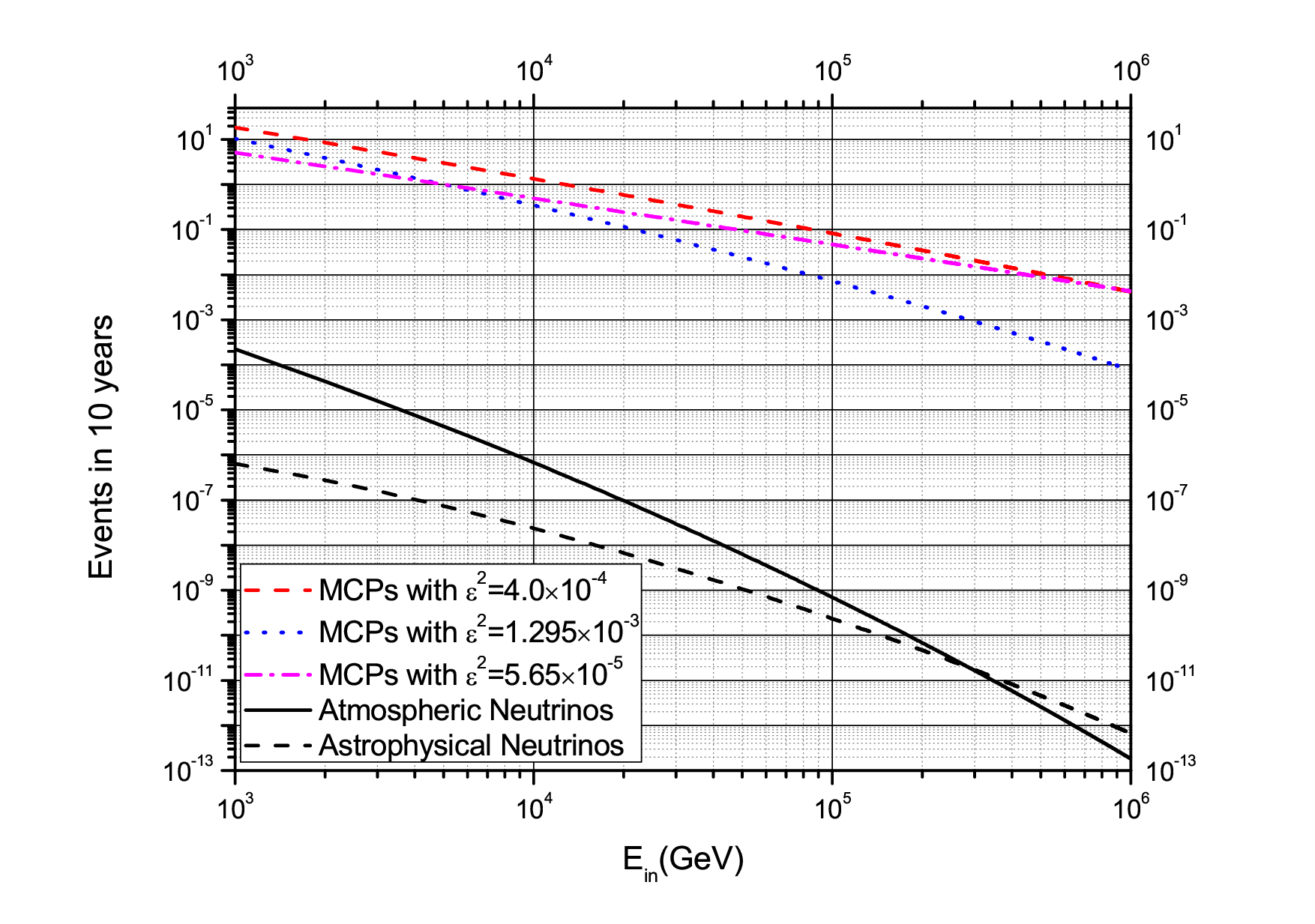}
 \caption{Distributions of expected MCPs with $\tau_{\phi}$ = $10^{18}$ s and $\epsilon^2=5.65\times10^{-5}$, $4\times10^{-4}$, $1.295\times10^{-3}$ and astrophysical and atmospheric neutrinos. Their energy bins are 100 GeV.}
 \label{fig:E_bin}
\end{figure}

The distributions and expected event counts of both MCPs and neutrinos were evaluated over the energy range of 1 TeV to 1 PeV, assuming 10 years of IceCube data. Figure~\ref{fig:E_bin} illustrates these distributions with an energy bin width of 100 GeV. Compared to MCPs with $\epsilon^2 = 1.295 \times 10^{-3}$ and $\tau_{\phi} = 10^{18}$ s (blue dotted line), the number of neutrino events per energy bin is at least five orders of magnitude lower above 5 TeV (With $m_{\phi}\ge$ 10 TeV as mentioned in Section 1, MCPs have energies above 5 TeV). As shown in Fig.~\ref{fig:E_bin}, the dominant background in this measurement comes from atmospheric neutrinos below 200 TeV, while astrophysical neutrinos prevail above approximately 400 TeV.
\par
\begin{figure}
 \centering
 \includegraphics[width=0.9\textwidth]{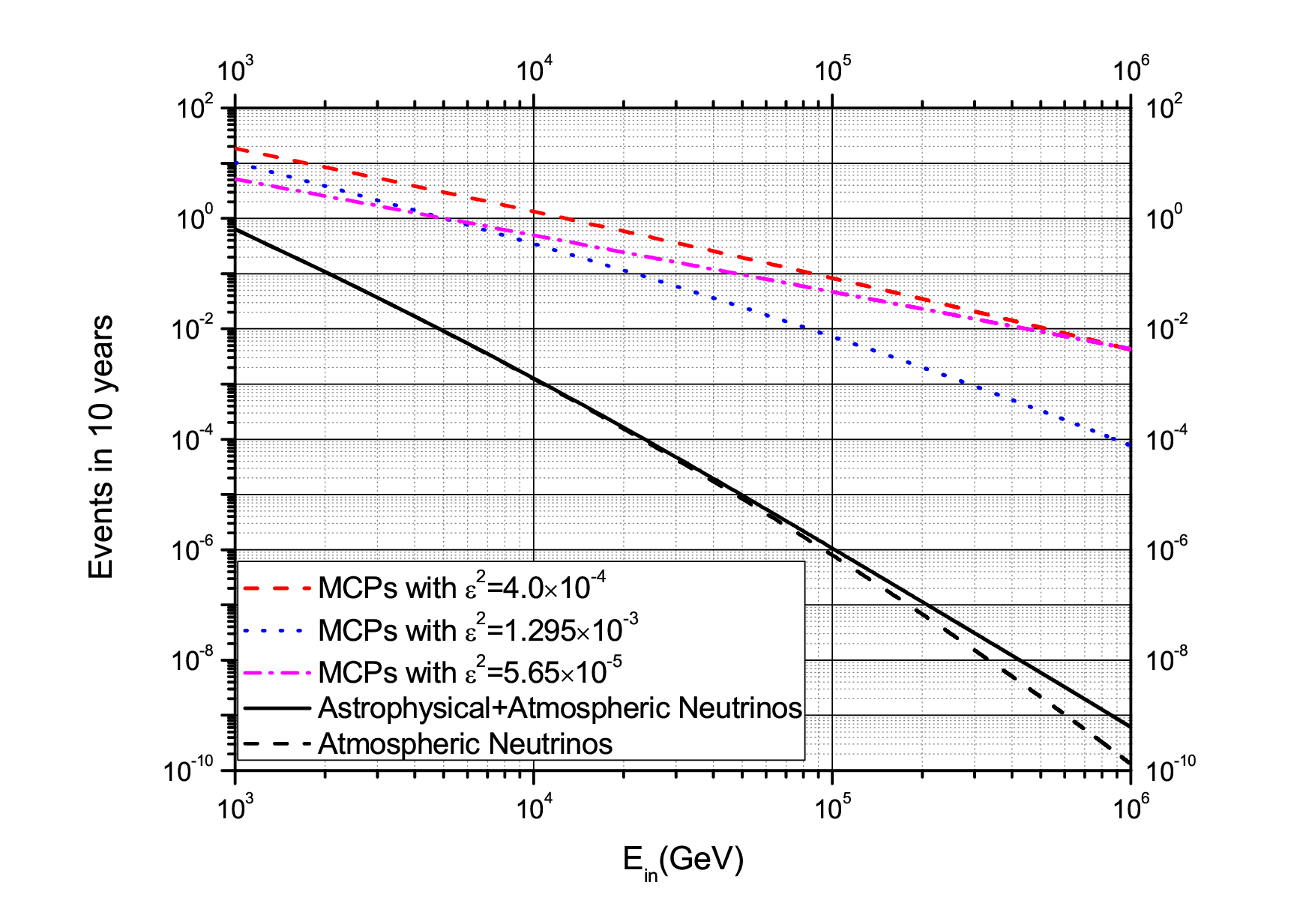}
 \caption{With the different $\epsilon^2$ (= $5.65\times10^{-5}$, $4\times10^{-4}$ and $1.295\times10^{-3}$) and $\tau_{\phi}$ = $10^{18}$, the numbers of expected MCPs were evaluated assuming 10 years of IceCube data, respectively. The evaluation of numbers of expected neutrinos was performed by integrating over the region caused by one standard energy and median angular uncertainties.}
 \label{fig:event_1e18}
\end{figure}
\begin{figure}
 \centering
 \includegraphics[width=0.9\textwidth]{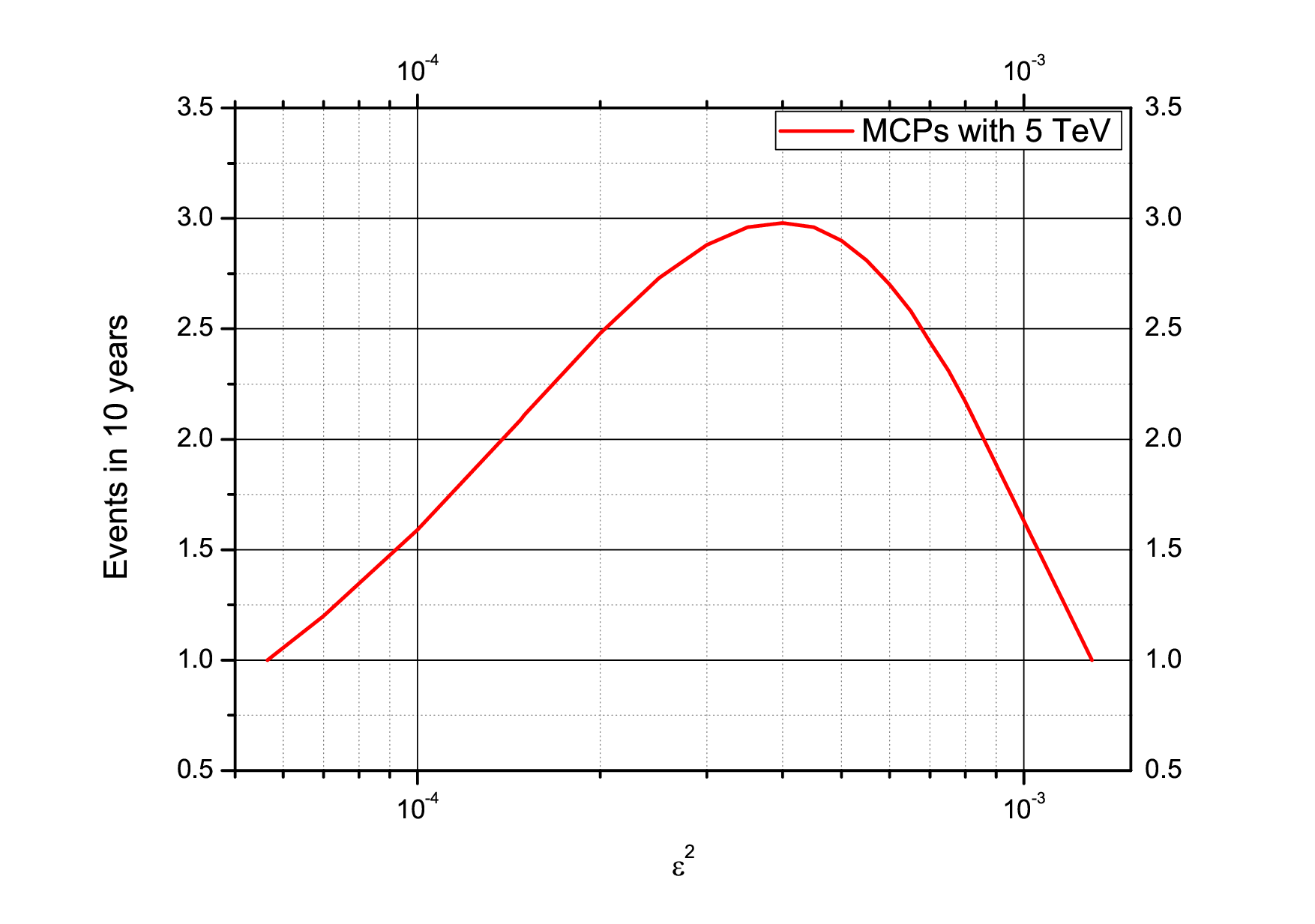}
 \caption{With the different $\epsilon^2$, the numbers of expected MCPs were evaluated at 5 TeV assuming 10 years of IceCube data, respectively. The number of
expected MCPs would reach a maximum value at IceCube.}
 \label{fig:N_5TeV}
\end{figure}

The total expected number of neutrino events (black solid line) is presented in Fig.~\ref{fig:event_1e18}, obtained by integrating over the regions defined by the energy and angular uncertainties described earlier. The black dashed line indicates the expected number of atmospheric neutrinos. This figure demonstrates that the neutrino background is negligible in the energy range of interest--for instance, the expected number of neutrinos is approximately 0.01 at 5 TeV. In contrast, the expected number of MCP events with $\epsilon^2 = 4 \times 10^{-4}$ and $\tau_{\phi} = 10^{18}$ s reaches about 19 at 5 TeV and 1 at 13 TeV, as shown by the red dashed line in Fig.~\ref{fig:event_1e18}. The same figure also displays expected event rates for MCPs with $\epsilon^2 = 5.65 \times 10^{-5}$ (magenta dash-dotted line) and $\epsilon^2 = 1.295 \times 10^{-3}$ (blue dashed line), which could be detected at about 5 TeV for $\tau_{\phi} = 10^{18}$ s. Furthermore, as shown in Fig.~\ref{fig:N_5TeV}, the expected MCP count reaches a maximum at IceCube for $\epsilon^2 \approx 4 \times 10^{-4}$.
\par
\begin{figure}
 \centering
 \includegraphics[width=0.9\textwidth]{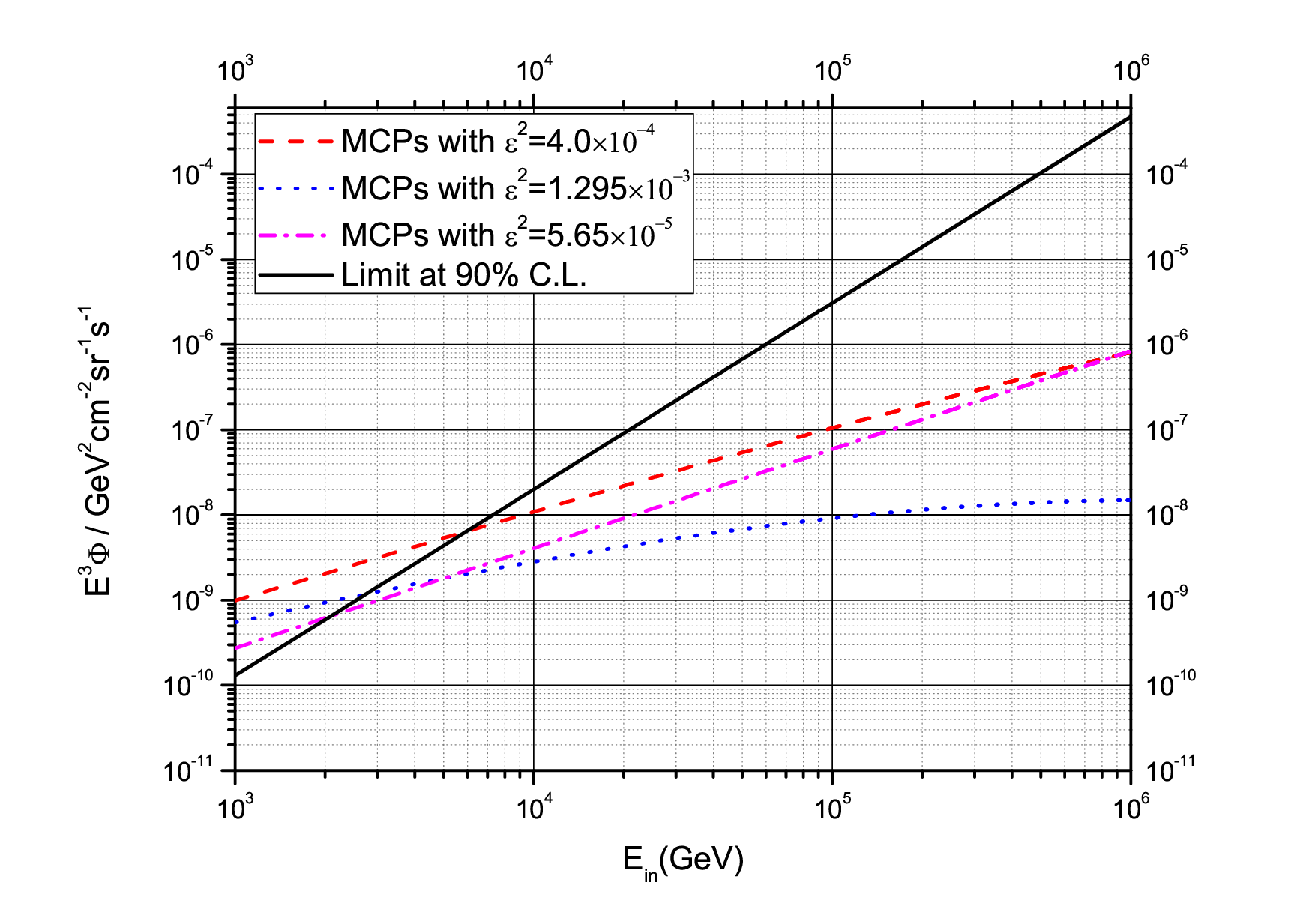}
 \caption{With the different $\epsilon^2$ (= $5.65\times10^{-5}$, $4\times10^{-4}$ and $1.295\times10^{-3}$) and $\tau_{\phi}$ = $10^{18}$, the fluxes of expected MCPs were estimated at IceCube, respectively. Assuming no observation at IceCube in 10 years, the upper limit at 90\% C.L. was also computed.}
 \label{fig:flux_1e18}
\end{figure}

A recent all-sky search for transient astrophysical neutrino emission using 10 years of IceCube cascade data \cite{icecube2024} reported no significant neutrino signal from the Earth's core. Given the difficulty in distinguishing MCP and neutrino signals at IceCube, it is reasonable to assume that no MCP events originating from $\phi$ decay in the Earth's core were observed over the 10-year period. The corresponding upper limit on the MCP flux at 90\% confidence level was computed using the Feldman--Cousins method \cite{FC} (black solid line in Fig.~\ref{fig:flux_1e18}). Figure~\ref{fig:flux_1e18} also shows expected MCP fluxes for $\epsilon^2 = 2 \times 10^{-5}$ (blue dotted line), $9 \times 10^{-4}$ (red dashed line), and $4.5 \times 10^{-3}$ (magenta dash-dotted line). The derived limit excludes MCP fluxes with $\epsilon^2 = 1.295 \times 10^{-3}$, $4 \times 10^{-4}$, and $5.65 \times 10^{-5}$ below approximately 2.8 TeV, 6 TeV, and 2 TeV, respectively.
\section{Discussion and Conclusion}
With $\epsilon^2 = 4\times10^{-4}$, MCPs originating from the Earth's core can be detected in the energy range 6-13 TeV at IceCube for a heavy dark matter lifetime of $\tau_{\phi} = 10^{18}$ s. Similarly, with $\epsilon^2 = 5.65\times10^{-5}$ and $1.295\times10^{-3}$, such MCPs are also accessible to IceCube observations. Based on the results presented above, it is reasonable to conclude that MCPs can be directly detected at IceCube at energies around $\mathcal{O}(1\ \text{TeV})$ for the charge range $5.65\times10^{-5} \lesssim \epsilon^2 \lesssim 1.295\times10^{-3}$. Since these constraints rely on the assumptions outlined earlier, we encourage experimental collaborations such as IceCube to perform an unbiased analysis using their full dataset.
\par
The MCP flux $\Phi_{\text{MCP}}$ is approximately proportional to $1 / \tau_{\phi}$ (see Eq. (2.3)), so the results depend significantly on the heavy dark matter lifetime. For instance, if $\tau_{\phi} = 10^{19}$ s, the expected number of MCP events at IceCube would be roughly a factor of 7 lower than that for $\tau_{\phi} = 10^{18}$ s.
\par
\begin{figure}
 \centering
 \includegraphics[width=0.9\textwidth]{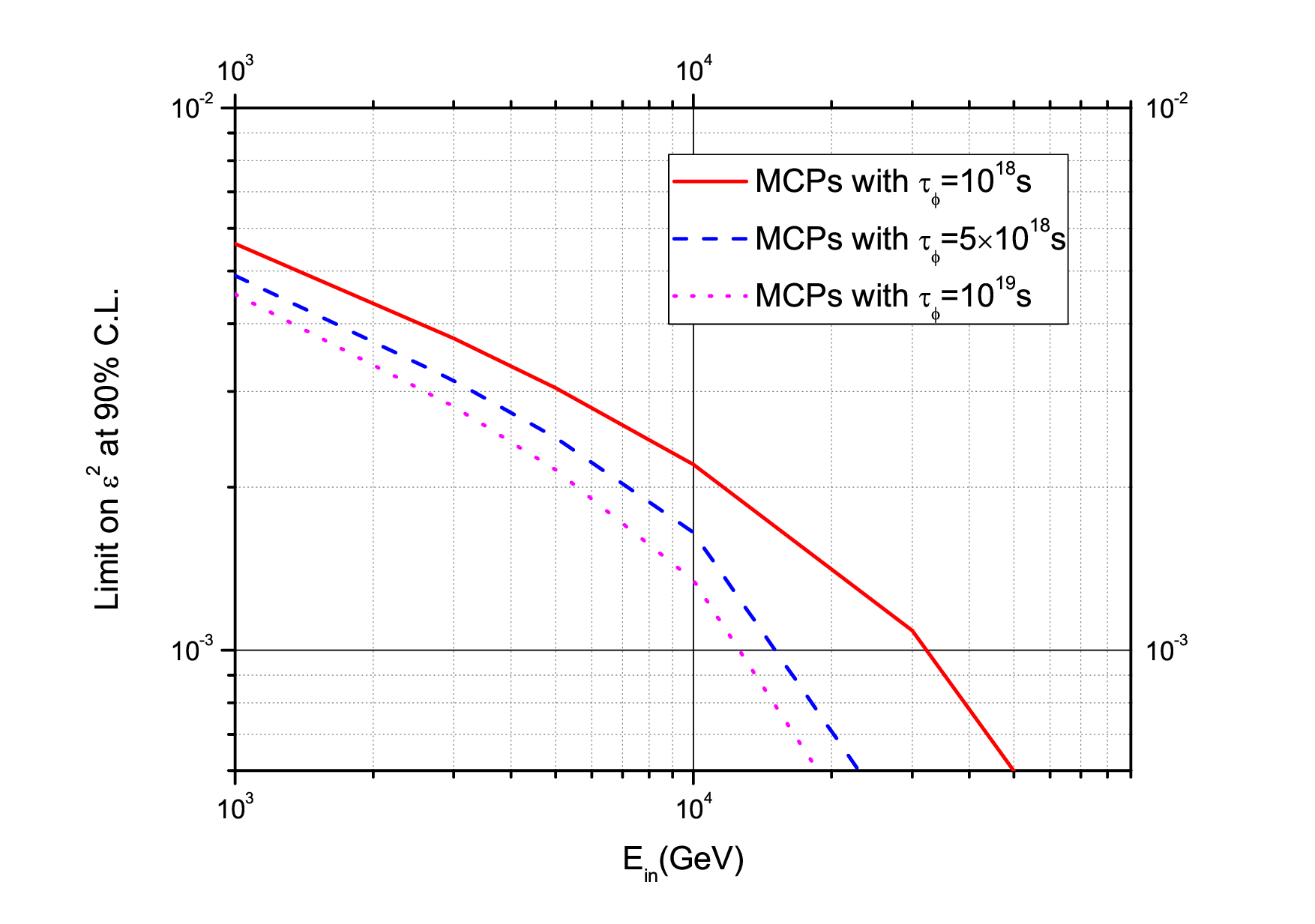}
 \caption{With the different $\tau_{\phi}$ (= $10^{18}$ s, $5\times10^{18}$ s and $10^{19}$ s), the upper limit on $\epsilon$ at 90\% C.L. was computed, respectively, assuming no observation at IceCube in 10 years.}
 \label{fig:uplimit_epsilon2}
\end{figure}

Using the Feldman--Cousins approach \cite{FC}, we also derive upper limits on $\epsilon^2$ at 90\% C.L. Figure~\ref{fig:uplimit_epsilon2} presents these limits for $\tau_{\phi} = 10^{18}$ s (red solid line), $5\times10^{18}$ s (blue dashed line), and $10^{19}$ s (magenta dotted line). For $m_{\phi} = 10\ \text{TeV}$, corresponding to an MCP energy of 5 TeV, the region $\epsilon^2 < 3.04\times10^{-3}$ (i.e., $\epsilon < 5.51\times10^{-2}$) is excluded for $\tau_{\phi} = 10^{18}$ s.
\par
\begin{figure}
 \centering
 \includegraphics[width=0.9\textwidth]{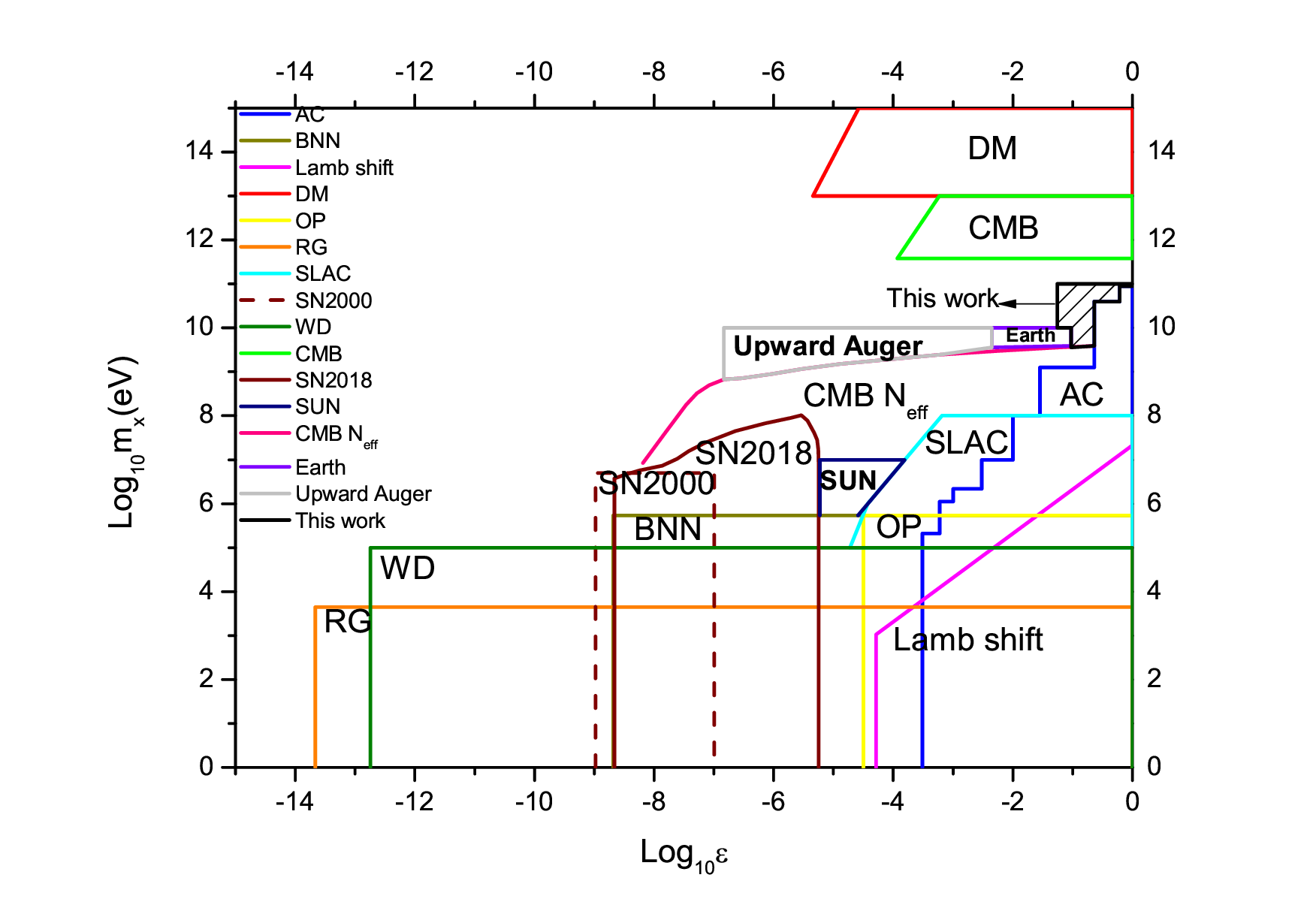}
 \caption{If $m_{\phi}$=2 TeV, a new region (shaded region) is ruled out in the $m_{\chi}$ vs. $\epsilon$ plane, when $\tau_{\phi}=10^{18}$ s. Meanwhile, the bounds from plasmon decay in red giants (RG)\cite{DHR}, plasmon decay in white dwarfs (WD)\cite{DHR}, cooling of the Supernova 1987A (SN2000\cite{DHR}, SN2018\cite{CM}), accelerator (AC)\cite{DCB} and fixed-target experiments (SLAC)\cite{SLAC}, the Tokyo search for the invisible decay of ortho-positronium (OP)\cite{OP}, the Lamb shift\cite{Lamb}, big bang nucleosynthesis (BBN)\cite{DHR}, cosmic microwave background (CMB)\cite{DGR}, dark matter searches (DM)\cite{JR}, measurement of MCPs from the sun's core\cite{SUN}, measurement of upward-going MCPs at Auger\cite{upauger} and the constraint on N$_{eff}$ at the CMB epoch by Planck\cite{VR}(N$_{eff}$ < 3.33\cite{Planck2018}) are also plotted on this figure.}
 \label{fig:epsilon_bound}
\end{figure}

As noted in Section 1, the MCP mass $m_{\chi}$ is taken to be below 100 GeV. Thus, in the $m_{\chi}$-$\epsilon$ plane, the region $\epsilon < 5.51\times10^{-2}$ is ruled out at 90\% C.L. for $m_{\chi} < 100\ \text{GeV}$, as illustrated in Fig.~\ref{fig:epsilon_bound}. For comparison, this figure also includes existing bounds on $\epsilon$ from cosmological and astrophysical observations \cite{CM,DHR,DGR,JR,VR}, accelerator and fixed-target experiments \cite{DCB,SLAC}, ortho-positronium decay studies \cite{OP}, Lamb shift measurements \cite{Lamb}, solar MCP searches \cite{SUN}, upward-going MCP limits from Auger \cite{upauger}, and the constraint on $N_{\text{eff}}$ from Planck CMB data \cite{VR,Planck2018} ($N_{\text{eff}} < 3.33$). Our analysis excludes a new region in the $m_{\chi}$-$\epsilon$ plane: $4\ \text{GeV} < m_{\chi} < 100\ \text{GeV}$ and $5.51\times10^{-2} < \epsilon < 0.612$, based on 10 years of IceCube data.
\par
In this dark matter scenario, the decay of $\phi$ into MCPs could lead to additional energy injection during the recombination epoch. However, since the $\phi$ lifetime exceeds the age of the Universe, the MCP relic density is given by
\begin{center}
\begin{equation}
\Omega_{MCPs}h^2=2(1-exp(-\displaystyle\frac{T_{re}}{\tau_{\phi}}))\Omega_{\phi}h^2
\approx2\displaystyle\frac{T_{re}}{\tau_{\phi}}\Omega_{\phi}h^2\lesssim 10^{-5}\Omega_{\phi}h^2
\end{equation}
\end{center}
where $T_{\text{re}}$ is the age of the Universe at recombination. Although MCPs can also be pair-produced via processes such as $e^+ + e^- \to \chi + \bar{\chi}$, potentially bringing them into thermal equilibrium with Standard Model particles in the early Universe, such processes are suppressed by $\epsilon^2$ and do not lead to significant energy injection during recombination. Our result is consistent with Ref.~\cite{DDRT}, which reported a strong cosmological constraint from Planck data, $\Omega_{\text{MCPs}} h^2 < 0.001$.
\section{Acknowledgements}
This work was supported by the National Natural Science Foundation
of China (NSFC) under the contract No. 11235006, the Science Fund of
Fujian University of Technology under the contracts No. GY-Z14061 and GY-Z13114 and the Natural Science Foundation of
Fujian Province in China under the contract No. 2015J01577.
\par

\newpage

\end{document}